# Epitaxial growth of $WO_3$ on r-plane sapphire by hot-filament metal-oxide deposition using $CO_2$ as a precursor

L. Souqui[1], L. J. Adams[1], A. Tsampanakis[1], D. Rehnlund Maibach[1,2], P. Eklund[1]

[1]Department of Chemistry – Ångström laboratory, Uppsala University, SE-751 21 Uppsala, Sweden

[2] Battery Division, Polestar R&D, Polestar Performance AB, SE-405 31 Göteborg, Sweden

**Abstract**

The deposition of epitaxial films of $WO_3$ by hot-filament metal-oxide deposition using $CO_2$ as a precursor was demonstrated. The films were obtained at a low growth rate of 0.25 nm/min. The optical properties of the film were assessed by variable-angle spectroscopic ellipsometry, indicating transparency across the visible spectrum and high absorption above 3 eV.

## Introduction

Tungsten trioxide, denoted $WO_x$ with $2 < x \leq 3$ is a versatile multifunctional material with a large compositional homogeneity ranges, and a variety of crystal structures.[1,2] The properties of interest range from semiconducting behavior with tunable bandgap, thermochromicity[3], electrochromicity [3,4], photochromicity[5], ferroelectricity[6], superconductivity, gas detection [7–9] and template for transition metal dichalcogenide growth[10]. Synthesis of $WO_x$ has been achieved by a range of physical[7,8,11–20] and chemical[6,9,10,21–26] vapor deposition techniques. One common approach to the synthesis of $WO_x$ is the hot-filament metal oxide deposition (HFMOD) method.[11,12,17–20] The method relies on the oxidation[27–30] of a hot tungsten wire by flowing $O_2$[17,18], $H_2O$[19] or air[11,12,17,20], and on the volatility of the oxides formed.[31]

Here, we instead use $CO_2$ as oxidizing species. $CO_2$ has long been known as an oxidizing agent of W, for example in reducing the lifetime of tungsten filaments in incandescent lamps[32,33] and in gas sensing[28]. Besides the advantage of upcycling a major greenhouse gas into valuable materials, $CO_2$ is a much milder oxidizer[28] than $O_2$ or $H_2O$. Furthermore, the formation of carbon monoxide (CO) as a byproduct of oxidation of the wires is expected to provide an additional means to tune the composition as it can act as a reducing agent for $WO_x$. This role of CO has been demonstrated in the case of the molybdenum oxides[34], and offers a means of engineering the composition and ultimately the properties of the deposited tungsten oxides. Thus, we here demonstrate epitaxial growth of orthorhombic (*o*-) or monoclinic (*m*-) $WO_3$ on r-plane sapphire by HFMOD using $CO_2$ as oxidizing gas.

## Experimental details

Tungsten oxide thin films were deposited onto r-plane sapphire substrates ($\alpha$-$Al_2O_3$(012), Alineason) in a hot-filament CVD (HFCVD) reactor (Blue Wave Semiconductors BWI 1000 HFCVD) by the reaction of tungsten (2 wires, 0.5-mm diameter, 8-cm long, 99.95% purity) with $CO_2$ (99.998% purity). The wires were placed at an average distance of 25 mm from the substrates.

The substrates were sonicated in acetone, isopropanol, and deionized water for 5 min each, then blow-dried using $N_2$ and loaded in an alumina crucible into a crucible furnace. The substrates were annealed in ambient air at 1000 °C for 60 min, with a temperature ramp-up of 5.4 °C/min and a passive cooldown. The substrates were stored in air no longer than three days.

Prior to loading into the HFCVD chamber, the substrates were sonicated once more in acetone and isopropanol for 5 min each, blow-dried using $N_2$. The base pressure of the system before deposition was < $4.0 \times 10^{-5}$ Pa ($3.0 \times 10^{-7}$ Torr). The depositions were run using 10 SCCM $CO_2$, resulting in a pressure of $7.2\times 10^{-2}$ Pa ($5.4 \times 10^{-4}$ Torr), at a substrate temperature of 450 °C, for 30-120 min, and with the substrate rotating at a speed of 12.5 rpm. Argon (Ar, 99.999%) was used at the same partial pressure of $7.2\times 10^{-2}$ Pa as a substitute to $CO_2$ to dismiss the role of impurities ($O_2$, residual $H_2O$) as potential oxidizers. The total current through the filaments was 35 A (17.5 A per filament). The corresponding filament temperature was estimated using the following equation:

$$T = \left(\frac{P_{Joule}}{\epsilon_W \, \sigma \, A} + T_0^4\right)^{\frac{1}{4}} = \left(\frac{\rho(T) I^2}{2\pi^2 r^3 \epsilon_W \sigma N} + T_0^4\right)^{\frac{1}{4}}, \quad (1)$$

where $T$ is the temperature of the wire, $T_0$ is the room temperature, $P_{Joule}$ is the power dissipated as heat by a resistor, $\epsilon_W$ is the emissivity of tungsten which we assumed constant and equal to 0.46 [35], $\sigma$ is the Stefan-Boltzmann constant, $A$ is the total emitting area, $I$ is the total current running through the filaments, $r$ is the radius of a filament and $N$ the number of identical filaments. The dependence of the resistivity of tungsten in the range 2000 to 3600 K is given by[36]:

$$\rho(T) = (-14.08 + 0.03515\,T) \times 10^{-8}[\,\Omega.\mathrm{m}] \tag{2}$$

which after 6 iterations of equations (1) and (2) converges to T = 2220 K (1947 °C).

X-ray diffraction (XRD) was measured using Cu Kα radiation. XRD 2θ/ω scans were recorded on a PANalytical Empyrean diffractometer with the integrated module iCore (Bragg-Brentano HD graded mirror with 0.8° acceptance angle, 0.03 rad Soller slits, 0.5° divergence slit at a sample distance of 96 mm) on the primary side and the integrated module dCore (17 mm anti-scatter slit, 0.02 rad Soller slits, 0.28° Parallel plate collimator) and PICel3D-Medipix3 detector on the secondary side. GID (Glancing incidence glancing exit diffraction) scans and X-ray pole figure were recorded on a X'Pert MRD diffractometer. A capillary lens with cross-slits was used on the primary side and a 0.18° parallel plate collimator and proportional detector were used on the secondary side, the cross-slits were set to 2×2 $mm^2$ for pole figure measurement and 8×2 $mm^2$ for GID measurements.

Variable-angle spectroscopic ellipsometry (VASE, RC2-XI Woollam) was used to evaluate the thickness and optical properties of the deposited films. The measurements were performed at incident angles of 50°, 55° and 60° of the range 0.73-5.9 eV. The data was modelled and fit using the CompleteEASE software (6.51).

**Results and discussion**

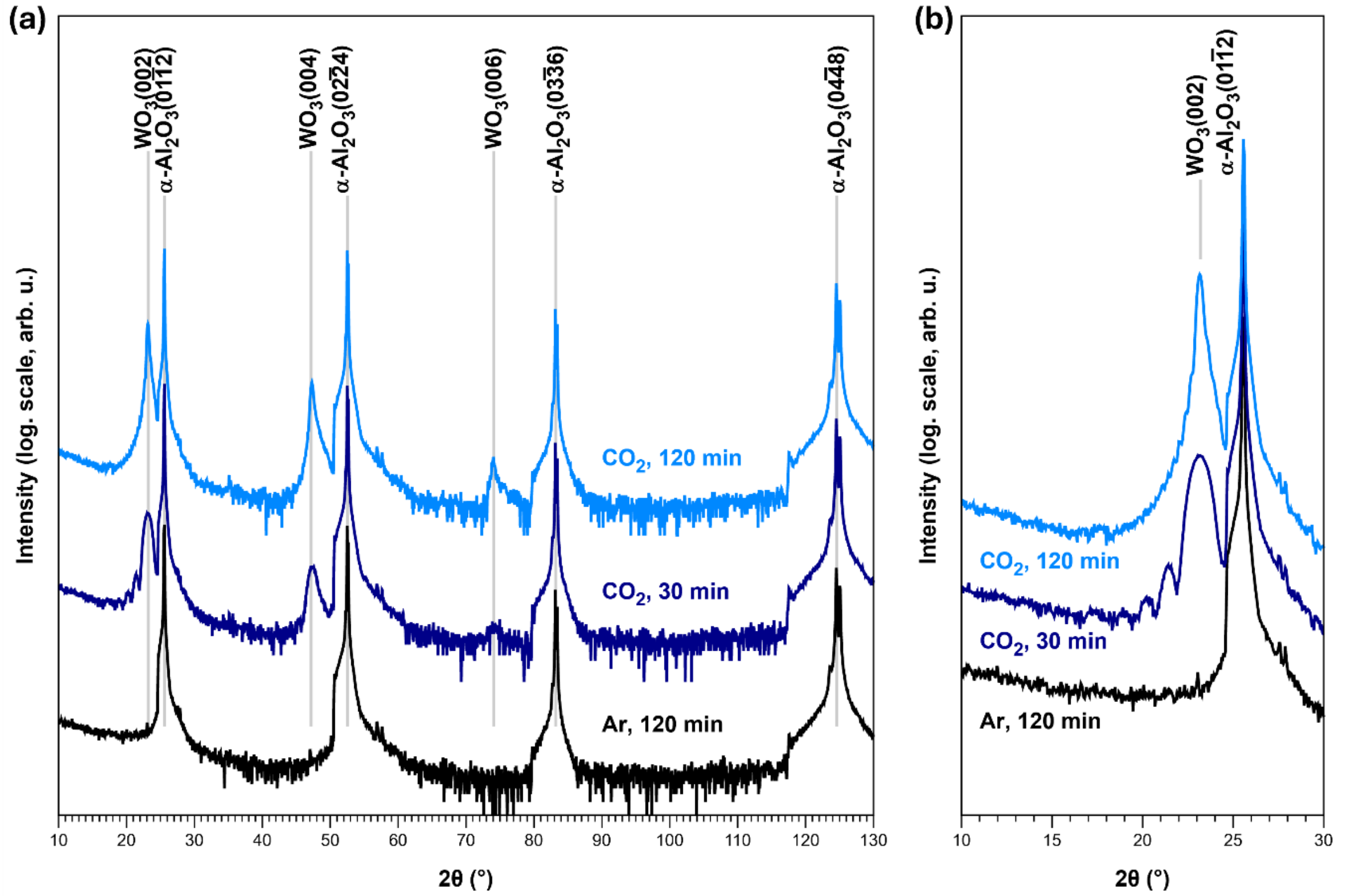

*Fig. 1: (a) Extended 2ϑ/ω scans of films deposited on sapphire with CO2 for 30 min, 120 min, and without CO2 (Ar). (b) Close-up around the tungsten oxide 002 peak*

Figure 1(a) shows 2θ/ω scans from films deposited with $CO_2$ for 30 min and 120 min, and for comparison films deposited with pure Ar. The XRD scan from the film deposited with $CO_2$ for 30 min exhibits distinct Laue fringes (Figure 1(b)), meaning that films of high crystalline quality are obtained. The fringes for the films deposited for 120 min are difficult to resolve because of their higher thickness of 30 nm, as measured by ellipsometry, corresponding to an average growth rate of 0.25 nm/min.

The 2θ/ω in Figure 1 and S1 showed a family of peak that is consistent with the {00*l*} planes of several phases of tungsten oxide, namely the monoclinic γ-$WO_3$, the orthorhombic β-$WO_3$, the high-temperature monoclinic phase, and the tetragonal phase.[2] No reflections besides the ones of the substrate are observed when substituting $CO_2$ by Ar with a similar impurity level. Indicating that $CO_2$ is the main oxidizing species in the process. The asymmetry of the peak is similar to the one reported by other authors for γ-$WO_3$[14] and tetragonal $WO_3$[8] on r-cut sapphire and may originate from the relaxation of the film.

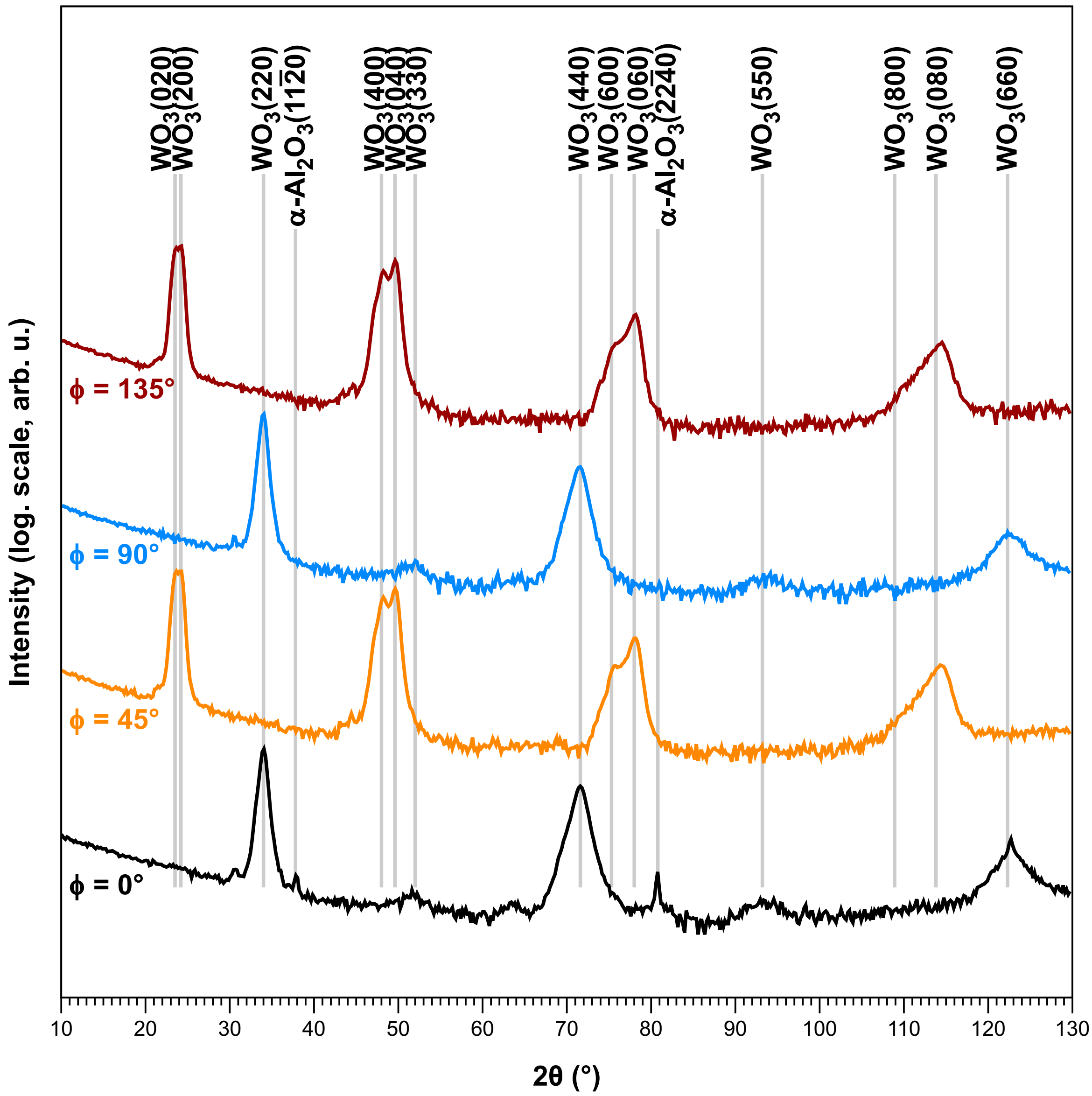

*Fig. 2: GID patterns of the 120 min-deposited sample taken along the [110] direction of sapphire, and rotated by φ = 45°, 90° and 135° and Ψ = 89.5°. Low intensity, non-indexed peaks and shoulders match with Kβ radiation.*

To get more insight into the symmetry of the phase, GID patterns and pole figures were recorded. The azimuth angle ϕ was set to zero for the scan aligned with the α-$Al_2O_3$$[110]$ (or α-$Al_2O_3$$[2\bar{1}\bar{1}0]$) direction. GID scans were recorded for ϕ = 0, 45, 90 and 135° (Figure 2), forming two sets of orthogonal scans and for Ψ = 89.5°, therefore corresponding to a 0.5° plane of incidence (assuming the sample perfectly flat). Only the {*hk*0} families of planes (with *h* or *k* non-zero) are observed. This confirms the out-of-plane relationship as

$$WO_3(001)// \alpha\text{-}Al_2O_3(01\bar{1}2).$$

At ϕ = 0° and ϕ = 90°, only the {*hk*0} plane families such as *h* = *k* (*e.g.* {220}) are found, from which we deduce the in-plane relationship to

$$WO_3(110)// \alpha\text{-}Al_2O_3(11\bar{2}0).$$

These epitaxial relationships are the same as those reported for monoclinic or orthorhombic $WO_3$ synthesized with other methods[7,14,16] and unlike reported in[8,21] where {00*l*} is also found in-plane.

Even the 'forbidden' diffraction peaks from the {330} and {550} families of planes are detected. The presence of both {*h*00} and {0*k*0} with similar intensities for both ϕ = 45° and 135° excludes the formation of high symmetry phases such as cubic and tetragonal $WO_3$ while indicating a 90° rotational twinning. Distinguishing between orthorhombic β-$WO_3$ (Pbcn) and monoclinic γ-$WO_3$ ($P2_1/n$) is known to be problematic.[7,16] The *h*00 and 0*k*0 peak positions match better with the orthorhombic phase than the monoclinic one, one may argue that the influence of the substrate may affect in-plane interplanar distances.

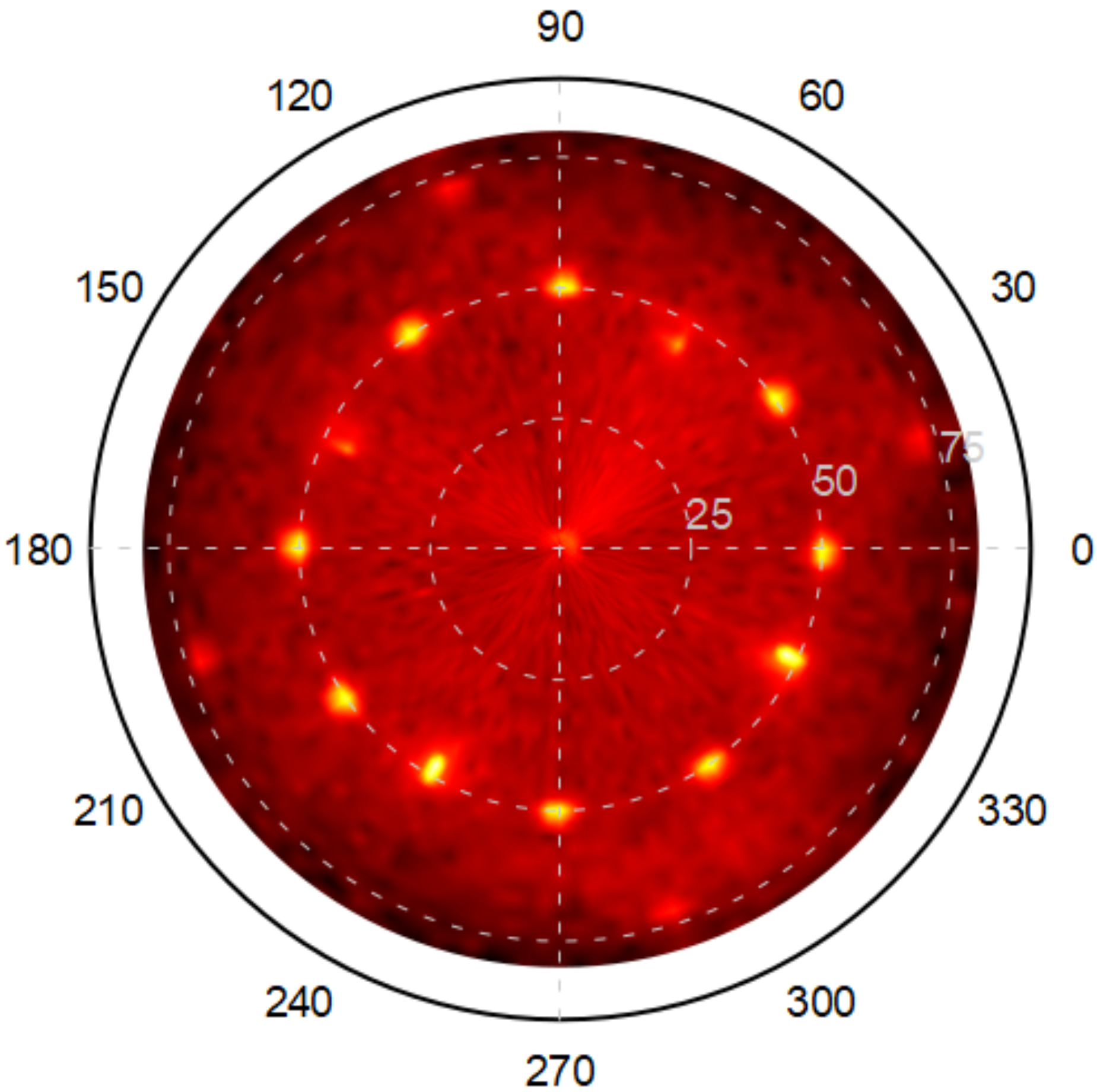

*Fig. 3: Pole figures recorded at 2ϑ = 35.68°.* The three set of circles correspond three groups of poles.

XRD pole figures taken at 2θ = 35.68° (Figure 3) confirm the epitaxial and twinned nature of the films. The poles in Figure 3 are distributed as follow: two sharp and weak poles at Ψ = 44.9°, two sharp and intense poles at Ψ = 47.7°, eight poles at Ψ = 50.0° and four broad and weak poles at Ψ = 72.2°. The sharp poles correspond to $Al_2O_3\{10\bar{1}4\}$ (2θ = 35.151, Ψ = 47°) and the tail of $Al_2O_3\{11\bar{2}0\}$ (2θ = 37.785°, Ψ = 43°). The eight poles at Ψ = 50.0° corresponds to $WO_3$, however, this can be interpreted either as the 122 poles of the orthorhombic phase, which appears as eight due to 90° rotational twinning (2×4 poles, the 212 reflections being extinct) or as the superposition of the 122, $12\bar{2}$, 212 and $21\bar{2}$ poles of the monoclinic phase, also resulting in eight poles. The monoclinic case would originate from four rotational twins due to its lower symmetry, however the broadening cause by the superposition is in the order of 1 to 2 degrees, which is lower than the step size of the measurement. Similarly, the poles at $\Psi$ = 72.2° are associated either to the 221 poles of the orthorhombic phase or the overlap of the 221 and $22\bar{1}$ poles of the monoclinic phase.

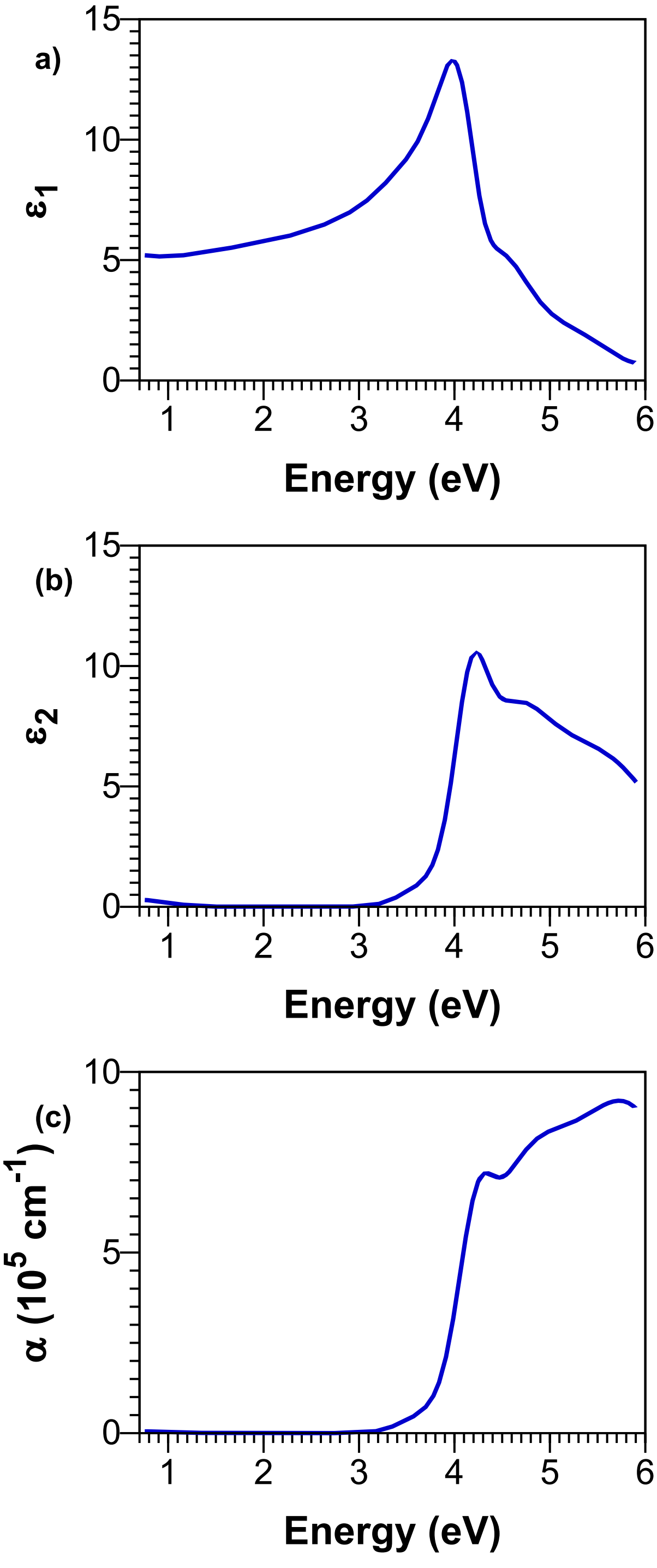

*Fig. 4: Optical properties of a 30-nm $WO_3$ thin film deposited on r-cut sapphire a) real part of the dielectric function, b) imaginary part of the dielectric function, and c) absorption coefficient. The feature at 4.5 eV is a fitting artifact.*

The real part of the complex dielectric function $\varepsilon_1$ obtained from VASE shows an increase from 5.15 at 1 eV to a maximum of about ≈13.3 at 4 eV (Figure 4). This matches well the one reported in a recent study[37] ($\varepsilon_1$ rising from 6 and reaching a maximum of 12 at 4 eV). The imaginary part of the dielectric function $\varepsilon_2$ and the absorption coefficient are low across most of the visible spectrum (with $\varepsilon_2 < 10^{-3}$ between 1.7 and 2.6 eV) up to around 3 eV where a sharp increase in absorption is observed. The latter value is consistent with reported value of 2.9 eV for the indirect band gap of stoichiometric $WO_3$.[38]

## Conclusion

We have demonstrated the epitaxial growth of monoclinic or orthorhombic $WO_3$ on r-cut sapphire by HFMOD using $CO_2$ as a precursor. The out-of-plane epitaxial relationship were $WO_3(001)//$ α-$Al_2O_3(01\bar{1}2)$ and the in-plane epitaxial relationships were either $WO_3(110)//$ α-$Al_2O_3(11\bar{2}0)$ or $WO_3(\bar{1}10)//$ α-$Al_2O_3(11\bar{2}0)$ due to 90°-rotational twining. The optical properties, obtained from spectroscopic ellipsometry measurements, were consistent with stoichiometric $WO_3$.

## Acknowledgments

We acknowledge funding from the Swedish Research Council VR under project no. 2025-03680, the Swedish Energy Agency under project no. P2025-03108, and the Knut and Alice Wallenberg foundation through the Wallenberg Academy Fellows program (KAW-2020.0196). We acknowledge Myfab Uppsala. Myfab is funded by the Swedish Research Council (2020-00207) as a national research infrastructure.